\documentclass[12pt]{article}
\usepackage{amsmath,amssymb}
\newcommand{\bos}{\boldsymbol}

\newcommand{\as}{a\!\!\!/}
\newcommand{\As}{A\!\!\!/}
\newcommand{\ks}{k\!\!\!/}
\newcommand{\ps}{p\!\!\!/}
\newcommand{\ets}{\eta\!\!\!/}

\newcommand{\ol}{\overline}

\date{}
\begin{document}
\baselineskip=18.6pt plus 0.2pt minus 0.1pt \makeatletter


\title{\vspace{-3cm}
\hfill\parbox{4cm}{\normalsize \emph{LPHEA 02-01}}\\
 \vspace{1cm}
{\bf Comment on Mott Scattering in Strong Laser Field.}}
 \vspace{2cm}

\author{Y. Attaourti \thanks{e-mail: attaourti@ucam.ac.ma},
 B. Manaut \thanks{e-mail: bmanaut@phea.ucam.ac.ma}\\
 {\it {\small Laboratoire de Physique des Hautes Energies et
d'Astrophysique,}}  \\{\it {\small Facult\'e des Sciences Semlalia,
Universit\'e Cadi Ayyad, Marrakech, BP : 2390, Maroc. }}}
\maketitle \setcounter{page}{1}
\begin{abstract}
The first differential cross section for Mott scattering of a
Dirac-Volkov electron is reviewed. The expression (26) derived by
Szymanowski et al. [Physical Review A {\bf 56}, 3846,(1997)] is
corrected. In particular, we disagree with the expression of
$\left(\frac{d\sigma}{d\Omega}\right)$ they obtained and we give
the exact coefficients multiplying the various Bessel functions
appearing in the scattering differential cross section.\\
\vspace{.04cm}\\
 PACS number(s): 34.80.Qb, 12.20.Ds
\end{abstract}

\section{Introduction}
\hspace{0.6cm}In a pioneering paper, Szymanowski et al. \cite{1}
have studied the Mott scattering process in a strong laser field.
The main purpose was to show that the modifications of the Mott
scattering differential cross section for the scattering of an
electron by the Coulomb potential of a nucleus in the presence of
a strong laser field, can yield interesting physical insights
concerning the importance and the signatures of the relativistic
effects. Their spin dependent relativistic description of Mott
scattering permits to distinguish between kinematics and
spin-orbit coupling effects. They have compared the results of a
calculation of the first Born differential cross section for the
Coulomb scattering of the Dirac-Volkov electrons dressed by a
circularly polarized laser field to the first Born cross section
for the Coulomb scattering of spinless Klein-Gordon particles and
also to the non relativistic Schrodinger-Volkov treatment. The
aim of this comment is to provide the correct expression for the
first-Born differential cross sections corresponding to the
Coulomb scattering of the Dirac-Volkov electrons. One the one
hand, We show that the terms proportional to $ sin(2\phi_0)$ are
missing in \cite{1}, where $\phi_0$ is the phase stemming from
the expression of the circularly polarized electromagnetic field.
The claim of \cite{1} that they vanish is not true. These terms do
not depend on the chosen description of the circular polarization
in cartesian components. On the other hand, We perform the
calculations with some details and throughout this work, we use
atomic units $(\hbar=e=m=1)$ where $m$ denotes the electron
mass.The abbreviation DCS stands for the differential cross
section.\\
\indent The organization of this paper is as follows : in Section
2, we establish the expression of the $S$-matrix transition
amplitude as well as the formal expression of scattering DCS. In
Section 3, we give a detailed account on the various trace
calculations and show that indeed there is a missing term
proportional to $\sin(2\phi_0)$ that is not equal to zero. This
term as well as a term proportional to $\cos(2\phi_0)$ contribute
to $\left(\frac{d\sigma}{d\Omega}\right)$ and multiply the product
$J_{s+1}(z)J_{s-1}(z)$ where $J_{s}(z)$ is an ordinary Bessel
function of argument $z$ and index $s$. The argument $z$
appearing in the above mentioned product will be defined later.
Then, we carry out the derivation of the correct expression of
the scattering DCS associated to the exchange of a given number
of laser photons. We end by a brief a conclusion in Section 4.
\section{The $S$-matrix element and the scattering differential cross section.}
Exact solutions of relativistic wave equations \cite{2} are very
difficult to obtain. However, in seminal paper, Volkov \cite{3}
obtained the formal solution of The Dirac equation for the
relativistic electron with 4-momentum $p^{\mu}$ inside a
classical monochromatic electromagnetic field $A^{\mu}$. These
solutions are called the relativistic Volkov states. The plane
wave electromagnetic field $A^{\mu}$ of 4-momentum $k^{\mu}$
 $(k_{\mu}k^{\mu}=k^2=0)$ depends only on the argument
 $\phi=k.x=k_{\mu}x^{\mu}$ and therefore $A^{\mu}$ is such that :
 \begin{equation}
A^{\mu}=A^{\mu}(k.x)=A^{\mu}(\phi) \label{1}
 \end{equation}
 The 4-vector $A^{\mu}$ satisfies the Lorentz gauge condition
 $\partial_{\mu}A^{\mu}=0$ or equivalently $k_{\mu}A^{\mu}=0$. The
 Dirac-Volkov equation in an external field $A_{\mu}$ is:
\begin{equation}
\left\{(\hat{p}-\frac{1}{c}A)^2-c^2-\frac{i}{2c}F_{\mu\nu}\sigma^{\mu\nu}\right\}\psi(x)=0
\label{2}
 \end{equation}
 where $F_{\mu\nu}$ is the electromagnetic field tensor
 $F_{\mu\nu}=\partial_{\mu}A_{\nu}-\partial_{\nu}A_{\mu}$ and
 $\sigma^{\mu\nu}=\frac{1}{2}[\gamma^{\mu},\gamma^{\nu}]$. The
 matrices $\gamma^{\mu}$ are the anticommuting Dirac matrices such
 that
 $\gamma^{\mu}\gamma^{\nu}+\gamma^{\nu}\gamma^{\mu}=2g^{\mu\nu}1_4$
 where $g^{\mu\nu}$ is the metric tensor
 $g^{\mu\nu}=diag(1,-1,-1,-1)$ and $1_4$ is the identity matrix in
 four dimensions. The solutions of Eq. (\ref{2}) are the
 relativistic Dirac-Volkov wave functions:
 \begin{equation}
\psi_p(x)=R(p)\frac{u(p,s)}{\sqrt{2p_0V}}e^{iS(x)} \label{3}
 \end{equation}
 where :
 \begin{equation}
R(p)=\exp{(\frac{\ks\As}{2c(k.p)})}=1+\frac{\ks\As}{2c(k.p)}\label{4}
 \end{equation}
 and the function $S(x)$ is given by :
\begin{equation}
S(x)=-p.x-\int_{0}^{k.x}\frac{1}{c(k.p)}\left[p.A(\xi)-\frac{1}{2c}A^2(\xi)\right]d\xi
\label{5}
 \end{equation}
 In Eq. (\ref{3}), $u(p,s)$ represents a Dirac bispinor which
 satisfies the free Dirac equation and is normalized according to
 $\ol{u}(p,s)u(p,s)=u^*(p,s)\gamma^0u(p,s)=2c^2$. We consider a
 circularly polarized field :
 \begin{equation}
A=a_1\cos(\phi)+a_2\sin(\phi)\label{6}
 \end{equation}
 where $\phi=k.x$. We choose $a_1^2=a_2^2=a^2=A^2$ and
 $a_1.a_2=a_2.a_1=0$. The Lorentz condition $k.A=0$ implies
 $a_1.k=a_2.k=0$. If one assumes that $A^{\mu}$ is
 quasi-periodic so that its time average is zero $\ol{A^{\mu}}=0$,
 then using the Gordon identity, the averaged 4-current is easily
 obtained :
\begin{equation}
\ol{j^{\mu}}=\frac{1}{p_0}\left\{p^{\mu}-\frac{1}{2c^2(k.p)}\ol{A^2}k^{\mu}\right\}\label{7}
 \end{equation}
 If one sets :
\begin{equation}
q^{\mu}=p^{\mu}-\frac{1}{2c^2(k.p)}\ol{A^2}k^{\mu}\label{8}
 \end{equation}
 this yields :
\begin{equation}
q.q=q^{\mu}q_{\mu}=m_*^2c^2\label{9}
 \end{equation}
 with :
\begin{equation}
m_*^2=1-\frac{\ol{A^2}}{c^4}\label{10}
 \end{equation}
 One often calls the averaged 4-momentum $q^{\mu}$ a
 quasi-impulsion. Note that $q^{\mu}=(Q/c,\bos{q})$. The quantity
 $m_*$ plays the role of an effective mass of the electron inside
 the electromagnetic field. For the study of the process of Mott
 scattering in presence of a laser field, we use the Dirac-Volkov
 wave functions \cite{3} normalized in the volume $V$:
 \begin{equation}
\psi_q(x)=R(p)\frac{u(p,s)}{\sqrt{2QV}}e^{iS(q,x)} \label{11}
 \end{equation}
 where :
\begin{equation}
R(p)=R(q)=1+\frac{1}{2c(k.p)}\ks\As=1+\frac{1}{2c(k.p)}(\ks.\as_1\cos(\phi)+\ks.\as_2\sin(\phi))\label{12}
 \end{equation}
 and :
 \begin{eqnarray}
S(q,x)&=&-q.x-\frac{(a_1.p)}{c(k.p)}\sin(\phi)+\frac{(a_2.p)}{c(k.p)}\cos(\phi)\nonumber\\
&=&-q.x-\frac{(a_1.q)}{c(k.q)}\sin(\phi)+\frac{(a_2.q)}{c(k.q)}\cos(\phi)\label{13}
 \end{eqnarray}
 We turn now to the calculation of the transition amplitude. The
 interaction of the dressed electrons with the central Coulomb
 field :
\begin{equation}
A^{\mu}=(\frac{Z}{|\bos{x}|},\bos{0})\label{14}
 \end{equation}
 is considered as a first-order perturbation. This is well
 justified if $Z\alpha \ll 1$, where $Z$ is the nuclear charge of
 the nucleus considered and $\alpha$ is the fine-structure
 constant. We evaluate the transition matrix element for the
 transition ($i\rightarrow f$) :
\begin{equation}
S_{fi}=\frac{iZ}{c}\int d^4x
\ol{\psi}_{qf}(x)\frac{\gamma^0}{|\bos{x}|}\psi_{qi}(x)\label{15}
 \end{equation}
 We first consider the quantity :
\begin{equation}
\ol{\psi}_{qf}(x)\frac{\gamma^0}{|\bos{x}|}\psi_{qi}(x)=\frac{1}{\sqrt{2Q_i
V }}\frac{1}{\sqrt{2Q_f V
}}\ol{u}(p_f,s_f)\ol{R}(p_f)\frac{\gamma^0}{|\bos{x}|}R(p_i)u(p_i,s_i)e^{-i(S(q_f,x)-S(q_i,x))}\label{16}
 \end{equation}
 We have :
\begin{equation}
e^{-i(S(q_f,x)-S(q_i,x))}=\exp[i(q_f-q_i).x-iz\sin(\phi-\phi_0)]\label{17}
 \end{equation}
 where $z$ is such that :
\begin{equation}
z=\sqrt{\alpha_1^2+\alpha_2^2}\label{18}
 \end{equation}
 whereas the quantities $\alpha_1$ and $\alpha_2$ are given by:
\begin{equation}
\alpha_1=\frac{(a_1.p_i)}{c(k.p_i)}-\frac{(a_1.p_f)}{c(k.p_f)},\,
\alpha_2=\frac{(a_2.p_i)}{c(k.p_i)}-\frac{(a_2.p_f)}{c(k.p_f)}
\label{19}
 \end{equation}
 and the phase $\phi_0$ is such that
 $\phi_0=\arccos(\alpha_1/z)=\arcsin(\alpha_2/z)=\arctan(\alpha_2/\alpha_1)$.
 It is important at this stage to perform intermediate
 calculations in order to reduce the numbers of $\gamma$ matrices
 that will appear when one calculates the scattering DCS. After
 some algebraic manipulations, one gets :
\begin{equation}
\ol{u}(p_f,s_f)\ol{R}(p_f)\gamma^0R(p_i)u(p_i,s_i)=\ol{u}(p_f,s_f)[C_0+C_1\cos(\phi)+C_2\sin(\phi)]u(p_i,s_i)
\label{20}
 \end{equation}
 where the three coefficients $C_0$, $C_1$ and $C_2$ are
 respectively given by :
\begin{eqnarray}
C_0&=&\gamma^0-2k_0a^2\ks c(p_i)c(p_f)\nonumber\\
C_1&=&c(p_i)\gamma^0\ks\as_1+c(p_f)\as_1\ks\gamma^0\nonumber\\
C_2&=&c(p_i)\gamma^0\ks\as_2+c(p_f)\as_2\ks\gamma^0\label{21}
 \end{eqnarray}
 with $c(p)=\frac{1}{2c(k.p)}$ and $k_0=k^0=\omega/c$. Therefore,
 the transition matrix element becomes :
\begin{eqnarray}
S_{fi}&=&\frac{iZ}{c}\int
d^4x\frac{1}{\sqrt{2Q_iV}}\frac{1}{\sqrt{2Q_fV}}
\ol{u}(p_f,s_f)[C_0+C_1\cos(\phi)+C_2\sin(\phi)]u(p_i,s_i)\nonumber\\
&\times&\exp[i(q_f-q_i).x-izsin(\phi-\phi_0)] \label{22}
 \end{eqnarray}
 We now invoke the well-known identities involving ordinary Bessel
 functions $J_s(z)$ :
\begin{eqnarray}
\left\{\begin{array}{c}
1\\
\cos(\phi)\\
\sin(\phi)\end{array}\right\}e^{-iz\sin(\phi-\phi_0)}=\sum_{s=-\infty}^{\infty}\left\{\begin{array}{c}
B_s\\
B_{1s}\\
B_{2s}\end{array}\right\}e^{-is\phi}\label{23}
\end{eqnarray}
with :
\begin{eqnarray}
\left\{\begin{array}{c}
B_s\\
B_{1s}\\
B_{2s}\end{array}\right\}=\left\{\begin{array}{c}
J_s(z)e^{is\phi_0}\\
(J_{s+1}(z)e^{i(s+1)\phi_0}+J_{s-1}(z)e^{i(s-1)\phi_0})/2\\
(J_{s+1}(z)e^{i(s+1)\phi_0}-J_{s-1}(z)e^{i(s-1)\phi_0})/2i\end{array}\right\}\label{24}
\end{eqnarray}
Evaluating the integrals over $x_0$ and $\bos{x}$ yields for
$S_{fi}$ :
\begin{equation}
S_{fi}=\frac{i4\pi
Z}{\sqrt{2Q_iV}\sqrt{2Q_fV}}\sum_{s=-\infty}^{\infty}\frac{2\pi\delta(Q_f-Q_i-s\omega)
}{|\bos{q}_f-\bos{q}_i-s\bos{k}|^2}M_{fi}^{(s)}\label{25}
\end{equation}
where the quantity $M_{fi}^{(s)}$ is defined by :
\begin{equation}
M_{fi}^{(s)}=\ol{u}(p_f,s_f)[C_0B_s+C_1B_{1s}+C_2B_{2s}]u(p_i,s_i)\label{26}
\end{equation}
To evaluate the DCS, we first evaluate the transition probability
per particle into final states within the range of momentum
$d\bos{q}_f$ :
\begin{eqnarray}
dW_{fi}&=&|S_{fi}|^2\frac{Vd\bos{q}_f}{(2\pi)^3}\nonumber\\
&=&\frac{(4\pi)^2Z^2}{2Q_iV.2Q_fV}\sum_{s=-\infty}^{\infty}\frac{T2\pi
\delta(Q_f-Q_i-sw)}{|\bos{q}_f-\bos{q}_i-s\bos{k}|^4}|M_{fi}^{(s)}|^2\frac{Vd\bos{q}_f}{(2\pi)^3}\label{27}
\end{eqnarray}
where we have used the rule of replacement :
\begin{eqnarray}
[2\pi\delta(Q_f-Q_i-sw)]^2& \rightarrow
&2\pi\delta(0)2\pi\delta(Q_f-Q_i-sw)\nonumber\\
&=&T2\pi\delta(Q_f-Q_i-sw)\label{28}
\end{eqnarray}
Next, we have for the transition probability per unit time :
\begin{equation}
dR_{fi}=\frac{dW_{fi}}{T}=\frac{(4\pi)^2Z^2}{2Q_iV.2Q_fV}\sum_{s=-\infty}^{\infty}\frac{2\pi
\delta(Q_f-Q_i-sw)}{|\bos{q}_f-\bos{q}_i-s\bos{k}|^4}|M_{fi}^{(s)}|^2\frac{Vd\bos{q}_f}{(2\pi)^3}\label{29}
\end{equation}
Dividing $dR_{fi}$ by the flux of incoming particles :
\begin{equation}
|\bos{J}^{inc}|=\frac{|\bos{q}_i|c^2}{Q_iV}\label{30}
\end{equation}
then using the relation
$|\bos{q}_f|d|\bos{q}_f|=\frac{1}{c^2}Q_fdQ_f$ and integrating
over the final energy, we get for the scattering DCS :
\begin{equation}
\frac{d\sigma}{d\Omega_f}=\left.\frac{Z^2}{c^4}\frac{|\bos{q}_f|}{|\bos{q}_i|}\sum_{s=\-\infty}^{\infty}\frac{|M_{fi}^{(s)}|^2}{|\bos{q}_f-\bos{q}_i-s\bos{k}|^4}\right|_{Q_f=Q_i+sw}=\left.\sum_{s=-\infty}^{\infty}\frac{d\sigma^{(s)}}{d\Omega_f}\right|_{Q_f=Q_i+sw}\label{31}
\end{equation}

where :
\begin{equation}
 \left.\frac{d\sigma^{(s)}}{d\Omega_f}\right|_{Q_f=Q_i+sw}=\left.\frac{Z^2}{c^4}\frac{|\bos{q}_f|}{|\bos{q}_i|}\frac{|M_{fi}^{(s)}|^2}{|\bos{q}_f-\bos{q}_i-s\bos{k}|^4}\right|_{Q_f=Q_i+sw}\label{32}
\end{equation}
The calculation is now reduced to the computation of traces of
$\gamma$ matrices. This is routinely done using Reduce \cite{4}.
We consider the unpolarized DCS. Therefore, the various
polarization states have the same probability and the actually
measured DCS is given by summing over the final polarization $s_f$
and averaging over the initial polarization $s_i$. Therefore, the
unpolarized DCS is formally given by :
\begin{equation}
 \frac{d\ol{\sigma}}{d\Omega_f}=\left.\sum_{s=-\infty}^{\infty}\frac{d\ol{\sigma}^{(s)}}{d\Omega_f}\right|_{Q_f=Q_i+sw}\label{33}
\end{equation}
where :
\begin{equation}
\left.\frac{d\ol{\sigma}^{(s)}}{d\Omega_f}\right|_{Q_f=Q_i+sw}=\left.\frac{Z^2}{c^4}\frac{|\bos{q}_f|}{|\bos{q}_i|}\frac{1}{|\bos{q}_f-\bos{q}_i-s\bos{k}|^4}\frac{1}{2}\sum_{s_i}\sum_{s_f}|M_{fi}^{(s)}|^2\right|_{Q_f=Q_i+sw}\label{34}
\end{equation}
\section{Trace calculations.}
Since the controversy is very acute and precise about the results
of the sum over the polarization
$\frac{1}{2}\sum_{s_i}\sum_{s_f}|M_{fi}^{(s)}|^2$, we devote a
whole section to the calculations of the various traces that
intervene in the formal expression of the unpolarized DCS given
by Eq. (\ref{34}). We have to calculate :
\begin{eqnarray}
\frac{1}{2}\sum_{s_i}\sum_{s_f}|M_{fi}^{(s)}|^2&=&\frac{1}{2}\sum_{s_i}\sum_{s_f}|\ol{u}(p_f,s_f)[C_0B_s+C_1B_{1s}+C_2B_{2s}]u(p_i,s_i)|^2\nonumber\\
&=&\frac{1}{2}\sum_{s_i}\sum_{s_f}|\ol{u}(p_f,s_f)\Lambda^{(s)}u(p_i,s_i)|^2\label{35}
\end{eqnarray}
with :
\begin{eqnarray}
\Lambda^{(s)}&=&[\gamma^0-2k_0a^2\ks c(p_i)c(p_f)]B_s\nonumber\\
&+&[c(p_i)\gamma^0\ks\as_1+c(p_f)\as_1\ks\gamma^0]B_{1s}\nonumber\\
&+&[c(p_i)\gamma^0\ks\as_2+c(p_f)\as_2\ks\gamma^0]B_{2s}\label{36}
\end{eqnarray}
using standard techniques of the $\gamma$ matrix algebra, one has
: \begin{equation}
\frac{1}{2}\sum_{s_i}\sum_{s_f}|M_{fi}^{(s)}|^2=\frac{1}{2}Tr\{(\ps_fc+c^2)\Lambda^{(s)}(\ps_ic+c^2)\ol{\Lambda}^{(s)}\}\label{37}
\end{equation}
with :
\begin{eqnarray}
\ol{\Lambda}^{(s)}&=&\gamma^0\Lambda^{(s)\dag}\gamma^0\nonumber\\
&=&[\gamma^0-2k_0a^2\ks c(p_i)c(p_f)]B^*_s\nonumber\\
&+&[c(p_i)\as_1\ks\gamma^0+c(p_f)\gamma^0\ks\as_1]B^*_{1s}\nonumber\\
&+&[c(p_i)\as_2\ks\gamma^0+c(p_f)\gamma^0\ks\as_2]B^*_{2s}\label{38}
\end{eqnarray}
There are nine main traces to be calculated. We write them
explicitly :
\begin{eqnarray}
&&\mathcal{M}_1=Tr\{(\ps_fc+c^2)C_0(\ps_ic+c^2)\ol{C}_0\}|B_s|^2\nonumber\\
&&\mathcal{M}_2=Tr\{(\ps_fc+c^2)C_0(\ps_ic+c^2)\ol{C}_1\}B_sB^*_{1s}\nonumber\\
&&\mathcal{M}_3=Tr\{(\ps_fc+c^2)C_0(\ps_ic+c^2)\ol{C}_2\}B_sB^*_{2s}\nonumber\\
&&\mathcal{M}_4=Tr\{(\ps_fc+c^2)C_1(\ps_ic+c^2)\ol{C}_0\}B^*_sB_{1s}\nonumber\\
&&\mathcal{M}_5=Tr\{(\ps_fc+c^2)C_1(\ps_ic+c^2)\ol{C}_1\}|B_{1s}|^2\label{39}\\
&&\mathcal{M}_6=Tr\{(\ps_fc+c^2)C_1(\ps_ic+c^2)\ol{C}_2\}B_{1s}B^*_{2s}\nonumber\\
&&\mathcal{M}_7=Tr\{(\ps_fc+c^2)C_2(\ps_ic+c^2)\ol{C}_0\}B_{2s}B^*_s\nonumber\\
&&\mathcal{M}_8=Tr\{(\ps_fc+c^2)C_2(\ps_ic+c^2)\ol{C}_1\}B^*_{1s}B_{2s}\nonumber\\
&&\mathcal{M}_9=Tr\{(\ps_fc+c^2)C_2(\ps_ic+c^2)\ol{C}_2\}|B_{2s}|^2\nonumber
\end{eqnarray}
To simplify the notations, we will drop the argument of the
various ordinary Bessel functions that appear. The diagonal terms
give rise to :
\begin{eqnarray}\left.\begin{array}{c}
\mathcal{M}_1\propto|B_s|^2=J_s^2\\
\mathcal{M}_5\propto|B_{1s}|^2=\frac{1}{4}(J^2_{s+1}+2J_{s+1}J_{s-1}\cos(2\phi_0)+J^2_{s-1})\\
\mathcal{M}_9\propto|B_{2s}|^2=\frac{1}{4}(J^2_{s+1}-2J_{s+1}J_{s-1}\cos(2\phi_0)+J^2_{s-1})\label{40}
\end{array}\right.
\end{eqnarray}
So, taking into account the fact that the traces multiplying
$|B_s|^2$, $|B_{1s}|^2$ and $|B_{2s}|^2$ are not zero, one
expects that terms proportional to $J_{s+1}J_{s-1}\cos(2\phi_0)$
will be present in the expression of the scattering DCS. The
first controversy between our work and the result of Szymanowski
et al \cite{1} concerns the traces $\mathcal{M}_6$
and$\mathcal{M}_8$. Since :
\begin{eqnarray}\left.\begin{array}{c}
\mathcal{M}_6\propto B_{1s}B^*_{2s}=\frac{i}{4}(J^2_{s+1}-2iJ_{s+1}J_{s-1}\sin(2\phi_0)-J^2_{s-1})\\
\mathcal{M}_8\propto B_{1s}^*
B_{2s}=\frac{-i}{4}(J^2_{s+1}+2iJ_{s+1}J_{s-1}\sin(2\phi_0)-J^2_{s-1})\label{41}
\end{array}\right.
\end{eqnarray}
and with little familiarity with the $\gamma$ matrix algebra, one
can see at once that if the corresponding traces are not zero
then the net contribution of $\mathcal{M}_6+\mathcal{M}_8$ will
contain a term proportional to $J_{s+1}J_{s-1}\sin(2\phi_0)$. We
shall demonstrate that in what follows. We have :
\begin{eqnarray}
\mathcal{M}_6&=&Tr\{(\ps_f c+c^2)C_1(\ps_i
c+c^2)\ol{C}_2\}B_{1s}B_{2s}^*\nonumber\\
&=&Tr\{(\ps_f
c+c^2)[c(p_i)\gamma^0\ks\as_1+c(p_f)\as_1\ks\gamma^0](\ps_i
c+c^2)\nonumber\\
&&[c(p_i)\as_2\ks\gamma^0+c(p_f)\gamma^0\ks\as_2]\}B_{1s}B_{2s}^*\label{42}
\end{eqnarray}
From now on, we define a 4-vector :
\begin{equation}
\eta^{\mu}=(1,0,0,0)\label{43}
\end{equation}
We can therefore write :
\begin{equation}
\gamma^0=\ets\label{44}
\end{equation}
Then, Eq. (41) becomes :
\begin{eqnarray}
\mathcal{M}_6&=&Tr\{(\ps_f c+c^2)C_1(\ps_i
c+c^2)\ol{C}_2\}B_{1s}B_{2s}^*\nonumber\\
&=&Tr\{(\ps_f c+c^2)[c(p_i)\ets\ks\as_1+c(p_f)\as_1\ks\ets](\ps_i
c+c^2)\nonumber\\
&&[c(p_i)\as_2\ks\ets+c(p_f)\ets\ks\as_2]\}B_{1s}B_{2s}^*\label{45}
\end{eqnarray}
In \cite{1}, the authors claim that the controversial
$\sin(2\phi_0)$ term disappear because it is proportional to terms
like $Tr\{(\ps_f c+c^2)\gamma^0\ks\as_1(\ps_i
c+c^2)\as_2\ks\gamma^0\}$. This term as well as $Tr\{(\ps_f
c+c^2)\as_1\ks\gamma^0(\ps_i c+c^2)\gamma^0\ks\as_2\}$ are indeed
zero but for $Tr\{(\ps_f c+c^2)\gamma^0\ks\as_1(\ps_i
c+c^2)\gamma^0\ks\as_2\}$ and $Tr\{(\ps_f
c+c^2)\as_1\ks\gamma^0(\ps_i c+c^2)\as_2\ks\gamma^0\}$ this is no
longer true. These terms are not zero and we give explicitly
their values :
\begin{eqnarray}
Tr\{(\ps_f c+c^2)\gamma^0\ks\as_1(\ps_i c+c^2)\gamma^0\ks\as_2\}
&=&Tr\{(\ps_f c+c^2)\as_1\ks\gamma^0(\ps_i
c+c^2)\as_2\ks\gamma^0\}\nonumber\\
&=&8w^2\{(a_1.p_f)(a_2.p_i)+(a_1.p_i)(a_2.p_f)\}\label{46}
\end{eqnarray}
In most case, the various traces are zero except when the cyclic
process of taking scalar products of pairs comes to products such
that :
\begin{eqnarray}\left.\begin{array}{c}
(k.\eta)(k.\eta)(a_1.p_i)(a_2.p_f)\\
(k.\eta)(k.\eta)(a_1.p_f)(a_2.p_i)\label{47}
\end{array}\right.
\end{eqnarray}
in which case, one has contributions proportional to
$w^2(a_1.p_i)(a_2.p_f)$ and $w^2(a_1.p_f)(a_2.p_i)$ respectively.
Explicitly, we give the result for $\mathcal{M}_6$ and
$\mathcal{M}_8$. One has :
\begin{eqnarray}
\mathcal{M}_6&=&\frac{w^2}{c^2}\{2\sin(2\phi_0)\left[\frac{(a_1.p_i)}{(k.p_i)}\frac{(a_2.p_f)}{(k.p_f)}+\frac{(a_2.p_i)}{(k.p_i)}\frac{(a_1.p_f)}{(k.p_f)}\right]J_{s+1}J_{s-1}\nonumber\\
&&+i[-\{(a_1.p_i)(a_2.p_f)+(a_1.p_f)(a_2.p_i)\}J^2_{s-1}\nonumber\\
&&+\{(a_1.p_i)(a_2.p_f)+(a_1.p_f)(a_2.p_i)\}J^2_{s+1}]\}\label{48}
\end{eqnarray}
while $\mathcal{M}_8$ is given by :
\begin{eqnarray}
\mathcal{M}_8&=&\frac{w^2}{c^2}\{2\sin(2\phi_0)\left[\frac{(a_1.p_i)}{(k.p_i)}\frac{(a_2.p_f)}{(k.p_f)}+\frac{(a_2.p_i)}{(k.p_i)}\frac{(a_1.p_f)}{(k.p_f)}\right]J_{s+1}J_{s-1}\nonumber\\
&&-i[-\{(a_1.p_i)(a_2.p_f)+(a_1.p_f)(a_2.p_i)\}J^2_{s-1}\nonumber\\
&&+\{(a_1.p_i)(a_2.p_f)+(a_1.p_f)(a_2.p_i)\}J^2_{s+1}]\}\label{49}
\end{eqnarray}
The fact that complex numbers appear in the expressions of
$\mathcal{M}_6$ and $\mathcal{M}_8$ is not surprising since the
former is the complex conjugate of the latter and their real sum
is such that :
\begin{eqnarray}
\mathcal{M}_6+\mathcal{M}_8&=&\frac{4w^2}{c^2}\sin(2\phi_0)\left[\frac{(a_1.p_i)}{(k.p_i)}\frac{(a_2.p_f)}{(k.p_f)}+\frac{(a_2.p_i)}{(k.p_i)}\frac{(a_1.p_f)}{(k.p_f)}\right]J_{s+1}J_{s-1}\label{50}
\end{eqnarray}
So, the first controversy is settled and there is indeed a term
containing $\sin(2\phi_0)$ in the expression of the scattering
cross section. To put an end to any further criticism, we give in
the Appendix the Reduce program we have written with the
necessary commentaries and observations so that anyone in the
scientific community having some knowledge of this powerful
symbolic computational software can easily try it and check our
results. Before writing our Reduce program, we have  extensively
studied the textbook by A. G. Grozin \cite{5} which is full of
worked examples in various fields of physics particularly in QED.
We give the final result for the unpolarized DCS for the Mott
scattering of a Dirac-Volkov electron :
\begin{eqnarray}\left.\begin{array}{c}
\frac{d\ol{\sigma}^{(s)}}{d\Omega_f}=\frac{Z^2}{c^2}\frac{|\bos{q}_f|}{|\bos{q}_i|}\frac{1}{|\bos{q}_f-\bos{q}_i-s\bos{k}|^4}\times\\
\frac{2}{c^2}\{J_s^2A+ \big(J^2_{s+1}+J^2_{s-1}\big)B
+\big(J_{s+1}J_{s-1}\big)C+J_s\big(J_{s-1}+J_{s+1}\big)D\}\label{51}
\end{array}\right.
\end{eqnarray}
where for notational simplicity we have dropped the argument $z$ in
the various ordinary Bessel functions. The coefficients $A$, $B$,
$C$ and $D$ are respectively given by :
\begin{eqnarray}
A&=&c^4-(q_f.q_i)c^2+2Q_fQ_i-\frac{a^2}{2}\left(\frac{(k.q_f)}{(k.q_i)}+\frac{(k.q_i)}{(k.q_f)}\right)+\frac{a^2\omega^2}{c^2(k.q_f)(k.q_i)}((q_f.q_i)-c^2)+\nonumber\\
&&\frac{(a^2)^2\omega^2}{c^4(k.q_f)(k.q_i)}+\frac{a^2\omega}{c^2}(Q_f-Q_i)\left(\frac{1}{(k.q_i)}-\frac{1}{(k.q_f)}\right)\label{52}
\end{eqnarray}
\begin{eqnarray}
B&=&-\frac{(a^2)^2\omega^2}{2c^4(k.q_f)(k.q_i)}+\frac{\omega^2}{2c^2}\left(\frac{(a_1.q_f)}{(k.q_f)}\frac{(a_1.q_i)}{(k.q_i)}+\frac{(a_2.q_f)}{(k.q_f)}\frac{(a_2.q_i)}{(k.q_i)}\right)-\frac{a^2}{2}+\nonumber\\
&&\frac{a^2}{4}(\frac{(k.q_f)}{(k.q_i)}+\frac{(k.q_i)}{(k.q_f)})-\frac{a^2\omega^2}{2c^2(k.q_f)(k.q_i)}\big((q_f.q_i)-c^2\big)+\nonumber\\
&&\frac{a^2\omega}{2c^2}(Q_f-Q_i)\left(\frac{1}{(k.q_f)}-\frac{1}{(k.q_i)}\right)\label{53}
\end{eqnarray}
\begin{eqnarray}
C&=&\frac{\omega^2}{c^2(k.q_f)(k.q_i)}\big(\cos(2\phi_0)\{(a_1.q_f)(a_1.q_i)-(a_2.q_f)(a_2.q_i)\}+\nonumber\\
&&\sin(2\phi_0)\{(a_1.q_f)(a_2.q_i)+(a_1.q_i)(a_2.q_f)\}\big)\label{54}
\end{eqnarray}
\begin{eqnarray}
D&=&\frac{c}{2}\left((\AA.q_i)+(\AA.q_f)\right)-\frac{c}{2}\left(\frac{(k.q_f)}{(k.q_i)}(\AA.q_i)+\frac{(k.q_i)}{(k.q_f)}(\AA.q_f)\right)+\nonumber\\
&&\frac{\omega}{c}\left(\frac{Q_i(\AA.q_f)}{(k.q_f)}+\frac{Q_f(\AA.q_i)}{(k.q_i)}\right)\label{55}
\end{eqnarray}
where $\AA=a_1\cos(\phi_0)+a_2\sin(\phi_0)$.
\subsection{Comparison of the coefficients.}
The argument about the missing term proportional to
$\sin(\phi_0)$ having been given a convincing explanation, we now
turn to other remarks along the same lines since there are indeed
other differences between our result and the result of \cite{1}.
We discuss now the difference occurring in our expression of the
coefficient $A$ and the corresponding one of \cite{1}. To make
the comparison easier we give explicitly the simple relations
between our coefficients and the corresponding coefficients of
\cite{1}. One has : $A(\cite{1})=A/c^2$, $B(\cite{1})=2B/c^2$,
$C(\cite{1})=2C/c^2$ and $D(\cite{1})=2D/c^2$. In their expression
multiplying the product $2J_n^2(\xi)$, the single term
$\frac{(a^6)^2w^2}{c^2(k.q)(k.q')}$ should come with a coefficient
$\frac{1}{2}$. In the appendix, we give a second Reduce program
that allows the comparison between the coefficient $A$ of
\cite{1} and the coefficient $A$ of this work. There are so many
differences between our result and the result they found for the
coefficient $B$ that we refer the reader to our main Reduce
program. The coefficient $C$ has already been discussed. As for
the coefficient $D$, we have found an expression that is linear
in the electromagnetic potential. In the appendix, we give a
third Reduce program. It is shown explicitly that if we ignore
the first term in the coefficient multiplying
$J_s(J_{s-1}+J_{s+1})$ given in \cite{1}, one easily gets the
result we have obtained. This term does not come from the passage
from the variables $(p,\tilde{p})$ to the variable
$(q,\tilde{q})$. The introduction of such 4-vector $\tilde{q}$ is
not useful, makes the calculations rather lengthy and gives rise
to complicated expressions. As a supplementary consistency check
of our procedure used in writing the main Reduce program, we have
reproduced the result of the DCS corresponding to the Compton
scattering in an intense electromagnetic field given by
Berestetzkii, Lifshitz and Pitaaevskii \cite{6}.
\section{Conclusion.}
In this comment, we derived the correct expression of the first
Born differential cross section for the scattering of the
Dirac-Volkov electron by a Coulomb potential of a nucleus in the
presence of a strong laser field. We have given the correct
relativistic generalization of the Bunkin and Fedorov treatment
\cite{7} that is valid for an arbitrary geometry. To prove that
our results are correct, we give the Reduce program to let the
scientific community judge their accuracy. We are adamant that
the core of the whole controversy stems from the fact that in
\cite{1}, the vector $\eta^{\mu}$ introduced in Eq. (\ref{43}) of
our work has not been properly dealt with while it is the common
method to use when a trace contains a $\gamma^0$ matrix. Any
standard QED textbook introduces this very elementary method.
\begin{center}
{\huge Appendix}
\end{center}
We give the main Reduce program that calculates the traces in Eq.
(\ref{37}). For this program to be readable, before every line,
we give an explanation of the different instructions. Some are
obvious, other are less straightforward and we also give the
number of the equation to which it refers in the text wherever it
is possible. In a Reduce program, a commentary is preceded by the
symbol $\%$. A Reduce instruction is not preceded by any symbol.
\begin{center}
{\huge The main program}
\end{center}
\% This program calculates the trace appearing in Eq. (\ref{37})
of the text. \\
\% The result must be multiplied by 4. Reduce calculates the
quarter of any trace.\\
\% This is well explained in the manual \cite{4}.\\
\% We first define the vector $p_f$, $p_i$, $a$, $a_1$, $a_2$,
$k$, $\eta$ (stands for nu), $q_i$, $q_f$. \\
 vector pfin, pin, aa, a1, a2, k, nu, qin, qfin; \\
\% The command mass associates the relevant scalar variable as a
mass with.\\
\% The corresponding vector. In the next instruction, cv stands
for the velocity of light.\\
 mass k=0, pfin=cv, pin=cv;\\
\% The command mshell put a particle 'on the mass shell'.\\
\% A substitution $<$vector  variable$>$.$<$vector
variable$>$=$<$mass$>$**2 is set up.\\
 mshell k, pfin, pin;\\
 \% The results of the above instruction are : $k^2=0,\,p^2_f=c^2$
 and $p^2_i=c^2$.\\
on div;\\
 \% We define properties of the vector $\eta$ introduced in Eq. (\ref{43}). \\
 let nu.nu=1, nu.k=w/cv, k.nu=w/cv;\\
let nu.pfin=efin/cv, pfin.nu=efin/cv, nu.pin=ein/cv,
 pin.nu=ein/cv;\\
let nu.a1=0, a1.nu=0, nu.a2=0, a2.nu=0;\\
 \% We define the Maxwell gauge condition.\\
let a1.k=0, k.a1=0, a2.k=0, k.a2=0;\\
 \% We define the properties of the electromagnetic field
 potential.\\
 \% We cannot use the variable $a$ in our program. \\
 \% Reduce interprets it as the matrix $\gamma^5$.\\
let a1.a1=aa.aa, a2.a2=aa.aa, a1.a2=0, a2.a1=0;\\
 \% We define ($\ps c+c^2$). The variable 'l' denote the
 fermionic line. \\
 for all p let gp(p)=cv*g(l,p)+cv**2;\\
 \% We define the properties of the various quantities stemming
 from Eq. (\ref{24}).\\
\% impart denotes the imaginary part and repart denotes the real
part.\\
\% js, jsp1 and jsm1 denote respectively $J_s$, $J_{s+1}$ and
$J_{s-1}$.\\
let impart(js)=0, impart(jsp1)=0, impart(jsm1)=0;\\
let repart(js)=js, repart(jsp1)=jsp1, repart(jsm1)=jsm1;\\
let impart(s)=0, impart(phi0)=0; \\
let repart(s)=s, repart(phi0)=phi0;\\
\% We define the quantities of Eq. (\ref{24}).\\
bs:=js*exp(i*s*phi0);\\
b1s:=(jsp1*exp(i*(s+1)*phi0)+jsm1*exp(i*(s-1)*phi0))/2;\\
b2s:=(jsp1*exp(i*(s+1)*phi0)-jsm1*exp(i*(s-1)*phi0))/(2*i);\\
repart(bs):=js*cos(s*phi0); \\
impart(bs):=js*sin(s*phi0);\\
repart(b1s):=(jsp1*cos((s+1)*phi0)+jsm1*cos((s-1)*phi0))/2;\\
impart(b1s):=(jsp1*sin((s+1)*phi0)+jsm1*sin((s-1)*phi0))/2;\\
repart(b2s):=(jsp1*sin((s+1)*phi0)-jsm1*sin((s-1)*phi0))/2;\\
impart(b2s):=(jsm1*cos((s-1)*phi0)-jsp1*cos((s+1)*phi0))/2;\\
\% We ask Reduce not to perform the various traces for the time
being. \\
 nospur l;\\
\% We define the various products appearing in Eq. (\ref{37}).\\
t1:=gp(pfin);\\
t2:=(g(l,nu)-2*c*cp*(aa.aa)*(k.nu)*g(l,k))*bs;\\
t3:=(c*g(l,nu)*g(l,k)*g(l,a1)+cp*g(l,a1)*g(l,k)*g(l,nu))*b1s;\\
t4:=(c*g(l,nu)*g(l,k)*g(l,a2)+cp*g(l,a2)*g(l,k)*g(l,nu))*b2s;\\
t5:=gp(pin);\\
t6:=(g(l,nu)-2*c*cp*(aa.aa)*(k.nu)*g(l,k))*conj(bs);\\
t7:=(cp*g(l,nu)*g(l,k)*g(l,a1)+c*g(l,a1)*g(l,k)*g(l,nu))*conj(b1s);\\
t8:=(cp*g(l,nu)*g(l,k)*g(l,a2)+c*g(l,a2)*g(l,k)*g(l,nu))*conj(b2s);\\
\% To obtain compact expressions, we define :\\
 for all fi, let cos(fi)+i*sin(fi)=exp(i*fi);\\
 for all fi, let cos(fi)-i*sin(fi)exp(-i*fi);\\
\% We explicitly define the coeficients $c(p_i)=c(q_i)$ and
$c(p_f)=c(q_f)$.\\
 c:=1/(2*cv*(k.qin);\\
 cp:=1/(2*cv*(k.qfin);\\
\% We define the product $J_s(J_{s-1}+J_{s+1})$ of Bessel
functions.\\
 let js*jsm1+js*jsp1=jsom;\\
\% We define the relation between the 4-vector $p$ and $q$.\\
 let pin=qin+(aa.aa)*c*k/cv;\\
 let pfin=qfin+(aa.aa)*cp*k/cv;\\
\% Same definition  for the energy.\\
let ein=gqi+(aa.aa)*c*w/cv;\\
let efin=gqf+(aa.aa)*cp*w/cv;\\
 \% We load the package ASSIST
\cite{4} that simplifies the
calculations.\\
 load\_package assist; \\
\% We now ask Reduce to calculate the traces related to the
fermionic line 'l'.\\
spur l;\\
\% Each trace res$i$ is calculated separately corresponds to
the trace $\mathcal{M}_i$ of Eq. (\ref{39}).\\
res1:=t1*t2*t5*t6; \\
trigreduce res1;\\
res2:=t1*t2*t5*t7;\\
trigreduce res2;\\
 res3:=t1*t2*t5*t8; \\
trigreduce res3;\\
res4:=t1*t3*t5*t6;\\
trigreduce res4;\\
res5:=t1*t3*t5*t7; \\
 trigreduce res5;\\
 res6:=t1*t3*t5*t8;\\
 trigreduce res6;\\
 res7:=t1*t4*t5*t6; \\
 trigreduce res7;\\
 res8:=t1*t4*t5*t7;\\
 trigreduce res8;\\
 res9:=t1*t4*t5*t8;\\
 trigreduce res9;\\
\% The total trace is the sum of all these traces.\\
 restot:=res1+res2+res3+res4+res5+res6+res7+res8+res9;\\
 trigreduce restot;\\
\% The program is complete and Reduce takes seconds to give the
answer.\\
\indent The most controversial terms correspond to res6 and res8.
It is very easy to check that our claim is well founded and that
there is indeed a term proportional to $sin(2\phi_0)$ that is
missing in \cite{1}. We now turn to the Reduce program that shows
that the single term $\frac{(a^2)^2w^2}{c^2(k.q)(k.q')}$ should
come with a coefficient $\frac{1}{2}$
\begin{center}
{\huge The coefficient A.}
\end{center}
\indent This coefficient comes from the calculation of
$\mathcal{M}_1$ in Eq. (\ref{39}). The coefficient $A$ of
\cite{1}. is transformed so that no vector $\tilde{q}$ appears
anymore. The symbols qit and qft stand for $\tilde{q}_i$ and
$\tilde{q}_f$ respectively. The instruction f1:=c1 is the correct
instruction that allows to find exactly the same result as that
given in Eq. (\ref{52}).\\

\noindent   on div; \\
vector qi, qf, qit, qft, k;\\
a1:=c**2-qi.qf+2*gqi*gqf/c**2;\\
b1:=aa**2*(k.qft/k.qi+k.qit/k.qf)/(2*c**2);\\
c1:=aa**4*w**2/(2*c**6*(k.qi)*(k.qf));\\
d1:=1-aa**2*w**2/(c**4*(k.qi)*(k.qf));\\
e1:=-aa**2*(1-k.qft*k.qit/(k.qi*k.qf))/(2*c**2);\\
 \% Here, we use the fact that there is a factor $1/2$ in the above mentioned single term.\\
f1:=c1;\\
g1:=aa**4*w**2*(k.qf*k.qft+k.qi*k.qit+aa**2*w**2/c**4);\\
h1:=g1/(2*c**6*(k.qi)**2*(k.qf)**2);\\
k.qit:=-k.qi+2*w*gqi/c**2;\\
k.qft:=-k.qf+2*w*gqf/c**2;\\
res:=(a1+b1+c1)*d1+e1+f1+h1;\\
This program gives exactly our coefficient $A$ given in Eq.
(\ref{52})
\begin{center}
{\huge The coefficient $D$.}
\end{center}
\indent For the coefficient $D$ of \cite{1}, we give the Reduce
program omitting the first term that contains a factor that is
quadratic in a. Doing so, we find exactly the same expression as
that given
in Eq. (\ref{55}).\\
  on div;\\
  vector qi, qf, qit, qft, k, aron;\\
  term1:=0;\\
  term2:=aron.qi*k.qit/(c*k.qf)+aron.qi*k.qft/(c*k.qi);\\
  term3:=aron.qi/c+aron.qf/c;\\
  k.qit:=-k.qi+2*w*gqi/c**2;\\
  k.qft:=-k.qf+2*w*gqf/c**2;\\
  resd:=term1+term2+term3;\\
\indent We have given convincing arguments to support our
results. We also gave programs that will allow anyone to check
every stage of our reasoning and to reach the same conclusion :
there are indeed mistakes in \cite{1}.

\end{document}